# Phonon density of states in lanthanide-based nanocrystals

Z. H. Li[1,2], D. Hudry[3,*], R. Heid[1], A. H. Said[4], M. D. Le[5], R. Popescu[6], D. Gerthsen[6], M. Merz[1], K. W. Krämer[7], D. Busko[3], I. A. Howard[3,8], B. S. Richards[3,8], F. Weber[1,*]

[1] Institute for Quantum Materials and Technologies, Karlsruhe Institute of Technology, 76021 Karlsruhe, Germany
[2] School of Energy and Power Engineering, Huazhong University of Science and Technology, Wuhan 430074, China
[3] Institute of Microstructure Technology, Karlsruhe Institute of Technology, 76021 Karlsruhe, Germany
[4] Advanced Photon Source, Argonne National Laboratory, Lemont, Illinois 60439, USA
[5] ISIS Facility, Rutherford Appleton Laboratory, Chilton, Didcot, Oxfordshire OX11 0QX, United Kingdom
[6] Laboratory for Electron Microscopy, Karlsruhe Institute of Technology, 76131 Karlsruhe, Germany
7 Department of Chemistry and Biochemistry, University of Bern, 3012 Bern, Switzerland
[8] Light Technology Institute, Karlsruhe Institute of Technology, 76131 Karlsruhe, Germany

* corresponding authors

We report a combined inelastic neutron and X-ray scattering study of the phonon density of states of the nano- and microcrystalline lanthanide-based materials $NaY_{0.8}Yb_{0.18}Er_{0.02}F_4$ and $NaGd_{0.8}Yb_{0.18}Er_{0.02}F_4$. While large (20 nm) nanocrystals display the same vibrational spectra as their microcrystalline counterparts, we find an enhanced phonon density of states at low energies, $E \leq 15$ meV, in ultra-small (5 nm) $NaGd_{0.8}Yb_{0.18}Er_{0.02}F_4$ nanocrystals which we assign to an increased relative spectral weight of surface phonon modes. Based on our observations for ultra-small nanocrystals, we rationalize that an increase of the phonon density of states in large nanocrystals due to surface phonons is too small to be observed in the current measurements. The experimental approach described in this report constitutes the first step toward the rationalization of size effects on the modification of the absolute upconversion quantum yield of upconverting nanocrystals.

# I. Introduction

Lanthanide (Ln)-based nanocrystals (NCs) are an important class of luminescent nanomaterials that can exhibit a wide-range of optical properties (upconversion, downconversion, and downshifting) [1-3], which are of major interest for potential technological applications [4-7]. In these materials, the luminescence arises from Ln dopants whose photophysical properties are intimately linked to intrinsic material characteristics. Among these material characteristics, the phonon density of states (PDOS) plays an important role in luminescence efficiency and dynamics as it governs the non-radiative relaxation processes between closely spaced electronic energy levels (both intra- and inter-ion). For instance, the multi-phonon relaxation pathways of an excited level of a given emitting center can be modified depending on the available phonon energies [Fig. 1(a)]. Additionally, phonons play a key role in controlling energy transfer processes between different optical centers (generally defined as donors and acceptors), for which the corresponding energy levels are not always resonant [Fig. 1(b)]. In such a case, the energy mismatch can be compensated by the emission or absorption of one or more phonons by the host lattice.

Theoretical investigations indicate that, at the nanoscale regime, the PDOS is not only modified from a continuous to a discrete distribution but also exhibits an energy threshold, which shifts to higher energy for smaller NCs [8]. For energies smaller than the threshold, all phonon modes are cut off. In the case of Ln-based NCs, cutoff energies ranging from 8 $cm^{-1}$ (1 meV) up to 30 $cm^{-1}$ (3.7 meV) as well as discrete phonon modes up to 200 $cm^{-1}$ (24.8 meV) have been deduced by several groups for a 2.5-20 nm size regime based on PDOS calculations [9-11]. The phonon confinement in Ln-based NCs has also been proposed to be responsible for various optical effects (anomalous hot bands, different spectral ratio, influence on luminescence efficiency, and modification of phonon-assisted relaxation and energy transfer pathways). However, the validity of these interpretations has been recently questioned [12]. While the impact of phonon properties is often discussed, the modification of the PDOS of Ln-based NCs as a function of the size is always derived from theoretical calculations while experimental measurements of the PDOS of NCs are still lacking.

Here, we report on the PDOS of Ln-based NCs $NaYF_4$ and $NaGdF_4$ doped with $Yb^{3+}$ (18 mol.%) and $Er^{3+}$ (2 mol.%) and their corresponding bulk-like counterparts. The PDOS were obtained from inelastic neutron scattering (INS) and non-resonant inelastic x-ray scattering (IXS) experiments. *Ab-initio* lattice dynamical calculations, which show a good agreement for microcrystalline samples, were used to identify the element-specific partial PDOS. Our results clearly reveal that i) large and slightly elongated NCs ($\approx$ 20 nm) exhibit the same vibrational spectra as their corresponding microcrystalline counterparts, and ii) ultra-small ($\approx$ 5 nm) NCs are characterized by an increase of the low-energy PDOS but do not indicate a phonon confinement gap in the accessible energy range E $\geq$ 3 meV. This is in agreement with another recent report [12] that indicated that a number of Ln luminescence effects which had been assigned to the depletion of phonon modes at low energies [10,13-17] could require revision. Our measurements likely indicate the presence of surface phonon modes which have to be discussed with regard to their impact on the optical properties of Ln-based NCs.

## II. Experiment

The samples $\beta$-NaY$_{0.8}$Yb$_{0.18}$Er$_{0.02}$F$_4$ and $\beta$-NaGd$_{0.8}$Yb$_{0.18}$Er$_{0.02}$F$_4$ will be hereafter referred to simply as NaYF$_4$ and NaGdF$_4$, respectively. The doping by 18 mol% Yb$^{3+}$ and 2 mol% Er$^{3+}$ was optimized for the upconversion emission of the materials. Both crystallize in the hexagonal space group $P\bar{6}$ (no. 174) with $Z = 1.5$, i.e. Na$_{1.5}$Ln$_{1.5}$F$_6$ per unit cell (Fig. 2) [18,19]. This structure is characterized by crystallographic sites with half occupation of Na (red-white) as well as a site half-randomly occupied by 50% Na and 50% Ln (red-blue) [20]. The microcrystalline samples were prepared according to a method previously described by Krämer and co-workers [21]. Highly monodisperse and non-agglomerated core NCs were prepared by a non-aqueous method as reported by Hudry and co-workers [22]. Two different sizes were investigated: isotropic ultra-small NCs (5.0 $\pm$ 1.2 nm - the size standard deviation is given as 3$\sigma$) and slightly anisotropic large ones (long axis: 28.6 $\pm$ 3.6, short axis: 21.0 $\pm$ 3.0) NCs were obtained. While we were able to synthesize large NCs for both NaYF$_4$ and NaGdF$_4$, ultra-small NCs could be only obtained for NaGdF$_4$. Indeed, ultra-small hexagonal NaYF$_4$ NCs are difficult to obtain without co-doping with Gd. In our attempts to synthesize ultra-small NaYF$_4$ NCs, only the cubic phase was stabilized. Note that to prevent the agglomeration of individual NCs, oleic acid (a long alkyl chain carboxylic acid with the formula C$_{17}$H$_{33}$COOH) is used as a stabilizing agent. Thus, although analysed as dry powders, all NC samples contain oleate (i.e. the corresponding carboxylate of oleic acid) ligands, which passivate the surface of the NCs.

The IXS experiments were carried out at the HERIX spectrometer [23] of the 30-ID beamline at the Advanced Photon Source, Argonne National Laboratory. All measurements were performed at room temperature. The incident energy was 23.72 keV [24] and the horizontally scattered beam was analysed by a set of diced spherical silicon analysers (Reflection 12 12 12) [25]. The full-width-at-half-maximum (FWHM) of the energy and wave vector space resolution was about 1.7 meV and 0.066 Å⁻¹, respectively. In general, different $|\textbf{Q}|$ values were investigated with one of the nine equally spaced analyser/detector pairs and, hence, the results coming from different analysers had to be normalized for the efficiency of the individual analysers. The efficiencies of the analysers were obtained by measuring the scattering of a piece of plastic with each analyzer positioned at the same scattering angle. We investigated powdery microcrystalline and NC samples of $NaY_{0.8}Yb_{0.18}Er_{0.02}F_4$ and $Na\,Gd_{0.8}Yb_{0.18}Er_{0.02}F_4$ with different crystal sizes of 5 nm, 20 nm and > 1µm. Experiments were done with an unfocused beam size of 2 mm × 0.5 mm in order to sample a large sample volume providing a random distribution of crystalline orientations. All powdery samples (microcrystalline and NCs) were sandwiched in between two kapton foils taped to a 0.3 mm thick piece of stainless steel featuring a 5 mm × 3 mm central hole. The lattice constants of microcrystalline $NaYF_4$ are $a = b = 5.915$ Å and $c = 3.496$ Å of the hexagonal unit cell, space group $P\bar{6}$. The corresponding lattice constants of $NaGdF_4$ are $a = b = 6.021$ Å and $c = 3.585$ Å in the same unit cell as $NaYF_4$. Constant momentum scans were performed for energy transfers from −10 to 70 meV and the typical counting time was about 45 seconds per point. Different absolute values of the momentum transfer $|\textbf{Q}|$ were investigated from 3.98 Å⁻¹ to 7.04 Å⁻¹, corresponding to the scattering angles $19° \leq 2\Theta \leq 33.8°$.

INS experiments were performed at the MARI time-of-flight chopper spectrometer located at the ISIS neutron scattering facility, Rutherford Appleton Laboratory [26]. The microcrystalline $NaYF_4$ powder (3 grams in total) was introduced in a standard thin-walled aluminium container. The same empty container was used as a reference and the corresponding spectrum was subtracted from the raw data. Measurements were done at room temperature using the rep-rate multiplication mode providing incident energies $E_i = 75$ meV, 20 meV and 9 meV. Since the $|\textbf{Q}|$ coverage with $E_i = 9$ meV is small, we only analysed data with $E_i = 20$ meV and 75 meV. The neutron-weighted PDOS was deduced by standard procedure within the Mantid program [27].

## III. Theory

Lattice dynamics calculations reported in this paper were performed in the framework of density functional perturbation theory (DFPT) within the mixed basis pseudopotential method [28,29] for stoichiometric $NaYF_4$. Scalar-relativistic norm-conserving pseudopotentials of Vanderbilt-type were constructed for Na, Y, and F [30], treating the 4s and 4p semicore states of Y as valence states. The mixed-basis scheme uses a combination of local functions and plane waves for the representation of the valence states [28], which allows for an efficient treatment of the fairly deep norm-conserving pseudopotentials. Local basis functions of $p$ and $d$ type at Na sites, of $s$ and $p$ type at F sites and of $s$, $p$, $d$ type at Y sites were supplemented by plane waves up to a kinetic energy of 20 Ry. For the exchange correlation functional, the local-density approximation in the parametrization of Perdue-Wang [31] was applied. Brillouin zone integrations were performed by $k$-point sampling in conjunction with the standard smearing technique [32] employing a Gaussian broadening of 0.2 eV.

In the calculation, the structure with the partially occupied sites (see above and Fig. 2) was approximated by an ordered 1×1×2 superstructure doubling the cell along the $c$-axis. Hexagonal lattice parameters were $a = b = 5.915$ Å and $c = 3.496$ Å, i.e. experimental ones [33]. The atomic positions were relaxed within the $a$-$b$ plane, while the positions along $c$ were kept fixed to the experimentally observed values [33]. Hexagonal 6×6×8 meshes corresponding to 288 $k$-points in the full Brillouin zone were used for structural optimization as well as for the calculation of dynamical matrices on a 2×2×2 hexagonal mesh. Phonon frequencies and eigenvectors at arbitrary points in the Brillouin zone were then obtained by Fourier interpolation of these dynamical matrices.

## IV. Results

### a) Calculated phonon density of states

DFPT calculations were done for stoichiometric $NaYF_4$ only because of the well-known problems including $f$ electron states in this technique. The chemical substitution was not included since the necessary huge unit cell would have rendered lattice dynamical calculations impossible. Hence, results for lattice dynamical properties for both $NaY_{0.8}Yb_{0.18}Er_{0.02}F_4$ and $NaGd_{0.8}Yb_{0.18}Er_{0.02}F_4$ were obtained based on the calculations for stoichiometric $NaYF_4$ for which we adjusted the average atomic mass and scattering cross sections for the rare earth site according to the chemical substitution levels. In the following, we again write $NaYF_4$ and $NaGdF_4$ for simplicity.

The calculated generalized PDOS for $NaYF_4$ is shown in Figure 3(a) (solid line) along with the corresponding partial PDOS for Na, Y and F (broken lines) including a broadening of 1.7 meV which reflects the typical energy resolution of our IXS experiments. The neutron-weighted PDOS [Fig. 3(b)] was calculated as $\sum_k \frac{\sigma_k}{m_k} \cdot g_k$, where $g_k$ is the partial PDOS of element $k$ and $\sigma_k$ and $m_k$ are the corresponding neutron scattering cross section and atomic mass. Furthermore, we computed the x-ray weighted PDOS for $NaYF_4$ [Fig. 3(c)] by using the corresponding x-ray scattering cross sections.

While the neutron-weighted PDOS shows a similar energy dependence as the generalized PDOS, the x-ray weighted PDOS is much more dominated by vibrations of the rare-earth ions [green dashed line in Fig. 3(c)] because of their large number of electrons. Hence, INS will allow us to observe the PDOS in NaYF$_4$ over the full energy range of the one-phonon cross section. Complementarily, IXS yields a clear picture of the low energy phonons of the rare-earth ions. Figure 3(d) shows the full calculated momentum-energy IXS spectrum for NaYF$_4$. To this end we performed IXS phonon structure factor calculations on a regular three dimensional grid with spacing of $0.05 \times \frac{2\pi}{a}$, $0.05 \times \frac{2\pi}{b}$ and $0.05 \times \frac{2\pi}{c}$ along the three axes of the reciprocal unit cell. Individual phonons for a particular wave vector $\boldsymbol{Q}$ were simulated by resolution-limited peaks using the calculated structure factors to scale the peak amplitudes. Subsequently, phonon intensities for wavevectors with the same absolute size $|\boldsymbol{Q}|$ were averaged. This was done in bins of 0.06 Å$^{-1}$ in $|\boldsymbol{Q}|$ corresponding to the momentum resolution of our IXS measurements. Importantly, the calculated $|\boldsymbol{Q}|$-averaged IXS intensity [red dashed line in Fig. 3(c)] is practically identical to the x-ray weighted PDOS [solid black line in Fig. 3(c)]. Please note that the intensities shown in Figure 3(d) include corrections for the x-ray scattering cross section, the atomic form factor, the $|\boldsymbol{Q}|^2$ dependence of phonon intensities, the Bose factor at room temperature and the factor energy$^{-1}$. Hence, these intensities are directly comparable to background subtracted inelastic x-ray spectra [see Fig. 5(c) below]. More details on the determination of the averaged IXS intensity are given in Appendix A. Our result shows that IXS data averaged over such a broad range in momentum space can be directly compared to the x-ray weighted PDOS and, hence, will be presented as such in the following section IV.C on IXS results.

### b) Inelastic neutron scattering

Experimentally, INS is the standard method used to probe the PDOS. In our study however, the strong neutron absorption of Gd renders neutron scattering in NaGdF$_4$ practically impossible. Moreover, the NCs can only be prepared as highly monodisperse NCs in small quantities ($\approx 50 - 100$ mg), which is not sufficient for INS experiments. More importantly, even if problems regarding the quantity of NCs can be overcome by designing appropriate scale-up synthesis protocols, as-prepared NCs are stabilized by organic ligands (oleates) that introduce a large number of hydrogen atoms, thus producing a significant and problematic background. Consequently, INS experiments were only performed on microcrystalline NaYF$_4$ whereas IXS experiments were performed on microcrystalline and NC samples both for NaYF$_4$ and NaGdF$_4$. It is worth noting that the two techniques yield complementary results in that the PDOS probed by neutrons is largely dominated by light element (F) vibrations whereas the x-ray cross section naturally highlights the partial PDOS of heavy elements such as rare earths.

The neutron-weighted PDOS with $E_i$ = 20 meV (black circles) and 75 meV (orange squares) for microcrystalline NaYF$_4$ is compared to our calculations in Figure 4 [34]. The calculated one-phonon neutron-weighted PDOS (orange solid line) and the corresponding partial PDOS of the elements (dashed lines) are plotted on top of the two-phonon contribution (dash-dotted line), which is considered by convoluting the calculated one-phonon PDOS with itself. Here, the energy values of the calculation were scaled by a factor of 1.08 in order to match the high-energy cut-off observed by INS. The one-PDOS were scaled to the experimentally observed area whereas the two-PDOS was matched to the observed intensities at $E \geq$ 63 meV. The comparison (see Fig. 4) shows that our calculations yield a good description of the PDOS of NaYF$_4$.

### c) Inelastic x-ray scattering – bulk material

While the neutron cross sections are similar for the different atomic species within NaYF$_4$, the x-ray scattering cross section for a given atom is strongly correlated with the number of electrons. Therefore, the IXS results are dominated by scattering from the heavy atoms vibrating at low energies. In order to follow the PDOS as function of the size, IXS experiments were performed on microcrystalline and NC samples for both NaYF$_4$ and NaGdF$_4$.

IXS raw data were acquired over a large momentum range from 3.98 to 7.04 Å$^{-1}$, e.g. for microcrystalline NaYF$_4$ [Fig. 5(a)]. A typical scan for $|\boldsymbol{Q}|$ = 5.97 Å$^{-1}$ is shown in Figure 5(b) (black circles). The elastic scattering was approximated by the experimentally determined resolution function and a constant background (black solid line). In general, multi-phonon contributions are present as well. We discuss this in detail in Appendix B and conclude that such contributions are small and should not significantly affect our analysis of the low-energy phonons. In fact, the best match with the experimental data was achieved employing only a constant background for the IXS raw data as shown in Figure 5(b). The approximated function was subtracted from the raw data to obtain the inelastic scattering intensities [orange squares in Fig. 5(b)] for energies $E \geq$ 3 meV. The resulting inelastic spectra reveal an intense band of phonons around 9 meV [Fig. 5(c)]. Finally, the inelastic spectra were |Q|-averaged in analogy to calculations [see dashed red line in Fig. 3(c)] and, thus, represent the X-ray weighted PDOS (see discussion in section IV.a).

The investigated $|\boldsymbol{Q}|$ values as well as the data analysis is the same for all compounds investigated by us with IXS. Hence, we discuss in the following only properties of the thus obtained X-ray weighted PDOS, which is shown for microcrystalline NaYF$_4$ and NaGdF$_4$ in Figures 6(a) and (b), respectively. Here, the scattered intensities have been corrected for the phonon thermal occupation factor for x-ray energy loss scattering $n + 1$ where $n$ is the Bose factor $n = 1/(e^{E/k_B T} - 1)$ with the phonon energy $E$ and the temperature $T$. As discussed in section IV.a, we compare experimental result to the calculated x-ray weighted PDOS [solid lines in Figs. 6(a,b)]. The energy axis was scaled by 1.08 as deduced from the comparison of the calculations with the INS data (see Fig. 4). We find that the broad peaks at 10 - 15 meV for NaYF$_4$ [Fig. 6(a)] and NaGdF$_4$ [Fig. 6(b)] are well explained by the calculations. Thus, the analysis of both the neutron and x-ray experiments on microcrystalline samples shows good agreement with the same *ab-initio* lattice dynamical calculation. The two experimental probes are complementary in that the different cross sections for neutrons and x-ray highlight vibrations of F atoms at $E \geq 20$ meV and vibrations of the heavy rare-earth elements at $E \leq 20$ meV, respectively. This sensitivity of IXS to the scattering from the rare-earth elements makes it ideal to investigate the vibrational properties of the optically active atomic species in Ln-based NCs.

### d) Inelastic x-ray scattering – nanocrystalline material

In the case of NaGdF$_4$, both isotropic ultra-small (5.0 ± 1.2 nm) and slightly anisotropic large (long axis: 28.6 ± 3.6, short axis: 21.0 ± 3.0) NCs were obtained. The corresponding high-angle annular dark-field scanning transmission electron microscopy images and size distribution histograms are shown in Figure 7 and prove the high quality of the as-synthesized NCs. The synthesized NaYF$_4$ NCs display similar size distributions as the large anisotropic NaGdF$_4$ NCs.

The IXS data taken on NC samples were processed in the same manner as the data for the microcrystalline samples. The resulting X-ray weighted PDOS data sets are shown in Figure 8(a) and compared to their corresponding microcrystalline counterparts. A quantitative comparison of the signal strength for the various samples was not possible due to the different quantities of material sampled by the x-ray beam. Thus, the observed x-ray weighted PDOS were normalized to have the same area in the energy range $E \geq 20$ meV.

As shown in Figure 8(a), there is no detectable change of the vibrational properties between the microcrystalline materials (green squares) and the large slightly anisotropic NCs both for NaGdF$_4$ and NaYF$_4$ (orange circles). On the contrary, there is a clear increase of the X-ray weighted PDOS at low energies $E \leq 15$ meV in the case of ultra-small NaGdF$_4$ NCs [red triangles, Fig. 8(a)]. The differences between nano- and microcrystalline samples are highlighted for the low-energy range in Figures 8(b) and (c).

Before addressing the implications with regard to the physics of nanocrystals, we need to discuss possible extrinsic sources of additional scattering in NCs. One possibility is additional scattering due to the presence of the oleate ligands, which are not present in the microcrystalline samples. However, we do not expect a detectable impact of the oleate ligands, bonded at the surface of the NCs, for two reasons. First, the oleate ligands mostly contain light atoms (H and C) and the PDOS obtained from IXS experiments are mostly sensitive to heavy elements. Second, large anisotropic NCs are also stabilized by the exact same ligands (with a quantity 2 to 3 times smaller) but no change with regard to the PDOS of the microcrystalline samples (with no stabilizing ligands) is observed. Thus, one can exclude an increase of inelastic scattering due to ligands larger than the statistical error bar and the observed increase at low energies cannot be attributed to scattering by oleate ligands. Another possibility is that the multi-phonon contribution is different in NC samples compared to microcrystalline ones. Here, we show in the Appendix that multi-phonon scattering is weak in general and cannot explain detectable changes at low phonon energies of $E \leq 10$ meV.

A small shift of the PDOS in NCs can also originate from a change of the lattice constants, which we have investigated using lab-based x-ray diffraction, the results of which are described in detail in Appendix C and Figure 13. Refining the obtained patterns for microcrystalline and 5 nm $NaGdF_4$, we find an increase of the unit cell volume of about 1.3%. Using 1.5 as a typical value for the Grüneisen parameter, we would expect a general phonon softening of 1.9% because of the softer lattice in 5 nm $NaGdF_4$. We replot the x-ray weighted PDOS for microcrystalline $NaGdF_4$ in Figure 9 along with that of 5 nm NCs. However, the energy values of the latter are up-scaled by 1.9% in order to compensate the expected softening because of the larger unit cell. The comparison shows that the x-ray weighted PDOS of the nanocrystalline sample is still significantly larger at small energies. Hence, a general phonon softening because of the larger unit cell cannot explain our results.

Our results indicate that the additional spectral weight in the PDOS of ultra-small $NaGdF_4$ NCs peaks at energies of 4-6 meV [Fig. 8(b)] which lies well below the lowest-energy peak of microcrystalline $NaGdF_4$ [see Fig. 6(b)]. This could indicate that primarily low-energy acoustic phonons soften reflecting a reduction of the speed of sound in NCs. Generally, intense acoustic phonon branches emanate at scattering angles with strong Bragg scattering, e.g. at $|\boldsymbol{Q}| = 5.6$ Å$^{-1}$ [see Fig. 5(a)]. Thus, one would expect a particularly strong effect close to such momentum transfers. However, we find that the effect of increased spectral weight in the IXS data in ultra-small NaGdF$_4$ NCs is evenly distributed over a large region in momentum space. Therefore, softening of acoustic phonons is also unlikely the origin of our observation.

## V.    Discussion

Theory predicts two effects in NCs: A phonon confinement gap purely related to the size of the NCs should appear [8] and the PDOS of NCs should feature enhanced tails at the low- and high-energy limits [35]. Indeed, the latter effects have been observed by nuclear-resonant inelastic x-ray scattering (NRIXS) of $^{57}$Fe in several studies on transition-metal nanoparticles [36-39]. Non-resonant scattering techniques were recently applied to PbS (INS) [40] and PbTe NCs (IXS) [41]. The neutron scattering study revealed a strong increase of low-energy phonons (≤ 5 meV) in PbS NCs with diameters of ≤ 8.2 nm. On the other hand, the IXS study on PbTe NCs (1.7 nm – 2.6 nm) did not report an enhanced PDOS compared to the bulk material, although an independent assessment is difficult since no raw data are shown [41].

A phonon confinement gap has not yet been identified experimentally in NCs and we see no sign of a phonon state depletion in our inelastic spectra on NaGdF$_4$ down to phonon energies of 3 meV (Fig. 8) in agreement with a recent study of high-resolution emission spectra in 10-20 nm size NaYF$_4$ NCs [12]. Indeed, in their work van Hest and co-workers demonstrated that signatures in emission spectra, which were previously assigned to phonon confinement, are simply due to sample heating induced by the laser excitation source. These signatures vanish for low laser power and the emission spectra were identical to those obtained for microcrystalline samples.

Ortigoza and co-workers have theoretically identified softened phonon modes similar to our observations as primarily tidal and torsional modes of atoms in the outermost positions of the NCs close to or at the surface [35]. In a different study, *ab-initio* molecular simulations explained experimentally observed softened phonon energies in PbS nanoparticles via surface phonon modes as well [40]. The relative spectral weight of such surface modes should naturally increase with decreasing NC size since the share of low-coordinated surface atoms increases strongly (as particle size decreases, the ratio of surface atoms to inner atoms increases). In order to assess the increase of the volume of the surface region $V_{surface}$ compared to the total particle volume $V_{total}$, we consider a spherical particle with radius $r$ and a thickness $\Delta$ of the surface region. Hence, $r_{s-v} = \frac{V_{surface}}{V_{total}} = 1 - [(r - \Delta)/r]^3$. In the limit of a very thin surface region, $\Delta \ll r$, $r_{s-v}$ increases as the particle radius decreases, i.e. by factors of 50 and 200 for $r = 20$ nm and 5 nm, respectively, compared to a particle with 1-$\mu$m diameter. This increase of a factor of four between 20 nm and 5 nm NCs explains that we observe the enhancement of the PDOS only in 5 nm and not in 20 nm NaGdF$_4$ NCs because the maximum of the enhancement is only 2-3 times larger than the scatter in the data [Fig. 8(b)]. If we consider a thicker surface region, the difference becomes smaller, e.g., 3.9 for $\Delta$ = 1.0 Å. But even with $\Delta$ = 6 Å, i.e. the value of the largest of the lattice parameters, $r_{s-v}$ increases by a factor of 3.3 from 20 nm to 5 nm NCs and, thus, rationalizes that we do not see the enhanced PDOS in the larger NCs. From the discussion above we conclude that soft surface phonon modes are the most likely origin of the enhanced PDOS observed for ultra-small NaGdF$_4$ nanocrystals.

It has been shown that the surface of NCs plays a crucial part for the photon upconversion emission intensity via the surface quenching effect. According to Wang et al. [42], the strongly reduced emission intensity in NCs (Fig. 10) is due to an increased number of optically active sites that are located at or near from the surface and for which the excitation energy can be easily quenched because of surface defects, impurities and ligands. The authors [42] demonstrated that the emission intensity of NaGdF$_4$:Yb:Tm NCs can be increased by more than two orders of magnitude simply by coating the NCs with a protecting (optically inactive) shell, which is one of the most efficient methods to boost the upconversion efficiency.

One can argue that the enhanced PDOS observed for the ultra-small core NCs (no protecting shell) might exacerbate the surface quenching effect by facilitating the energy migration process toward surface quenching sites. The calculated increase of the relative volume of the surface region $r_{s-v}$ (factor of 40-50 depending on $\Delta$) is of similar magnitude to the observed decrease of the absolute quantum yield in 20 nm NaYF$_4$ compared to its microcrystalline counterpart (factor of 20) (Fig. 10). While the correlation between energy dissipation and surface phonons is intriguing we emphasize that experimentally we could not observe an increased spectral weight of surface phonon modes in 20 nm NCs (see Fig. 8). Consequently, it remains uncertain whether the surface phonons in Ln-based NCs can be considered as a main factor that could account for reduced upconversion efficiency.

## VI. Conclusion

In summary, we report an IXS scattering study of the vibrational properties in Ln-based NCs backed up by INS measurements and DFPT calculations for their corresponding microcrystalline counterparts. We demonstrate that IXS yields a good measure of the PDOS in the investigated materials and detect a clear increase of the low-energy PDOS but only for NCs as small as 5 nm which is most likely related to surface phonon modes. We show schematically that the reduced relative volume of the surface region in 20 nm NCs can explain the unchanged IXS spectra – compared to those of microcrystalline samples - but emphasize that more detailed experiments are needed to assess the role of surface phonon modes with respect to the optical properties of Ln-based NCs.

**Acknowledgement**

The authors would like to thank Daniel Biner (Univ. Bern) for his assistance with the synthesis of microcrystalline samples. The authors would like to thank the KNMF (Karlsruhe Nano Micro Facility) for TEM access. Z.L. was supported by the Helmholtz-OCPC Postdoc Program. M.M. was supported by KNMF.This research used resources of the Advanced Photon Source, a U.S. Department of Energy (DOE) Office of Science User Facility operated for the DOE Office of Science by Argonne National Laboratory under Contract No. DE-AC02-06CH11357. Experiments at the ISIS Neutron and Muon Source were supported by a beamtime allocation RB1990222 from the Science and Technology Facilities Council. This work was performed on the supercomputer ForHLR funded by the Ministry of Science, Research and the Arts Baden-Württemberg and by the Federal Ministry of Education and Research.

**Appendix A: Calculated IXS intensities**

Figure 3(d) shows the full calculated IXS spectrum for $NaYF_4$ over a wide range in phonon energy and momentum. Importantly, the |$\boldsymbol{Q}$|-averaged intensity derived from these data shows very good agreement with the properly weighted PDOS [red dashed and black solid lines in Figs. 3(c) and 11(e)(f)]. Hence, IXS data taken over the same energy momentum range can be directly compared to the calculated PDOS.

In the following, we describe the procedure to calculate data such as shown in Figure 3(d). Based on our ab-initio calculations we performed phonon structure factor calculations on a regular three dimensional grid with spacing of $0.05 \times \frac{2\pi}{a}$, $0.05 \times \frac{2\pi}{b}$ and $0.05 \times \frac{2\pi}{c}$ over all Brillouin zones with 3.98 Å$^{-1}$ ≤ |$\boldsymbol{Q}$| ≤ 7.04 Å$^{-1}$. For each fixed wave vector $\boldsymbol{Q}$ phonons were simulated by resolution-limited peaks using the calculated structure factors to scale the peak amplitudes. Subsequently, phonon intensities for wavevectors with the same absolute size |$\boldsymbol{Q}$| were averaged in bins of |$\boldsymbol{Q}$| = 0.06 Å$^{-1}$. Some exemplary data for wave vectors with |$\boldsymbol{Q}$| = 5.99 Å$^{-1}$ are shown in Figure 11(a) (broken lines) along with the corresponding intensity averaged over all $\boldsymbol{Q}$ values with this absolute size [solid line]. Results for four different |$\boldsymbol{Q}$| values are shown in Figure 11(b). The complete data set is given in Fig. 3(d). These results include the

|$\boldsymbol{Q}$| dependent x-ray scattering cross section, the Bose factor for room temperature and the factor energy$^{-1}$ in the phonon scattering intensities in order to simplify comparison with experimental data shown in Figure 5(c).

The data calculated for the 16 |$\boldsymbol{Q}$| values which we investigated experimentally [see vertical dotted lines Fig. 5(c)] were used to generate the |$\boldsymbol{Q}$|-averaged data sets. The individual (broken lines) and averaged data (solid lines) are shown for NaYF$_4$ and NaGdF$_4$ in Figures 11(c) and (d), respectively. Finally, |$\boldsymbol{Q}$|-averaged data are compared to the respective PDOS calculations including the same factors (|$\boldsymbol{Q}$| dependent x-ray scattering cross section, Bose factor, energy$^{-1}$) for NaYF$_4$ [Fig. 11(e)] and NaGdF$_4$ [Fig. 11(f)].

## Appendix B: Multi-phonon contribution to X-ray weighted PDOS

In comparison to INS on time-of-flight spectrometers, there are two factors which complicate the determination of a multi-phonon contribution to the experimentally observed PDOS with IXS: (1) An experimental determination the background (comparable to an empty-can measurement in INS) for all analyzer positions would be very time consuming and was not possible in the allocated IXS beam time. (2) The dominant scattering by the heavy atoms, i.e. Y/Gd/Yb/Er, yields a peak in the X-ray weighted PDOS at fairly low energies [see Fig. 6]. Correspondingly, the peak in the, e.g., two-phonon contribution appears at energies where the one PDOS is not yet zero.

From INS we know that the one PDOS extends to energies just above 60 meV (see Fig. 4). Hence, the IXS intensities at E ≥ 65 meV are due to an experimental background and/or multi-phonon contributions. We illustrate different scenarios in Figure 12. We note that for the data shown in Figure 12 only the resolution-limited elastic line without a constant background [see Fig. 5(b)] was subtracted from the raw IXS data which then were processed to obtain the X-ray weighted PDOS. Hence, we find finite values at E ≥ 65 meV, which we can assign to the experimental background or a multi-phonon contribution.

If we assign 90% of the signal at high energies to constant background we cannot achieve a good match of the one PDOS on top of the combined background with the experimental data [Figs. 12(a),(c)]. The agreement is improved when we consider only a constant background and no multi-phonon contribution [Figs. 12(b)(d)]. Hence, we estimate that two-phonon scattering contributes 5% or less of the signal at high energies and, thus, does not significantly contribute to the main peak in the PDOS of the investigated materials. In order to present a clear analysis, we decided to present our results considering only a constant experimental background for the raw IXS data as illustrated in Figure 5(b).

## Appendix C: XRD patterns of microcrystalline and ultra-small nanocrystalline NaGdF₄

X-ray powder diffraction patterns were obtained at room temperature (25 °C) in Bragg–Brentano geometry using a D2Phaser diffractometer from Bruker (30 kV – 10 mA) with a copper anticathode ($K_{\alpha 1}$ and $K_{\alpha 2}$) and are shown for micro- and 5 nm NaGdF₄ in Figure 13. The data were acquired with 2.5° primary and secondary Soller slits, a fixed divergence slit (either 1 mm or 0.2 mm for nanocrystals and microcrystals, respectively), a nickel $K_{beta}$ filter, and the Lynxeye detector.

A full structure refinement was done for microcrystalline NaGdF₄ using FULLPROF [43]. The corresponding lattice parameters, atomic positions, and occupation numbers, which are consistent with published data [18-20], are given within Figure 13. When fixing the occupation of the Gd, F1, and F2 sites to 100 %, we find a half-occupied Na1 site and a 1:1 occupational disorder between Na and rare earth atoms on the Na2 site. The 57 % occupation of this site with Gd results from the fact that only Na-Gd disorder was taken into account for the refinement and, consequently, reflects that Gd is partially (~ 20 %) replaced with the heavier rare earths Yb and Er. The Rietveld pattern fitting [44]  for the sample of ultra-small NaGdF₄ NCs was performed with fixed atomic positions and occupation numbers because of the broad peaks.

The refinements of the data reveal an increase of the lattice parameters in nanocrystalline NaGdF₄ (see table) corresponding to a volume change of 1.27%.

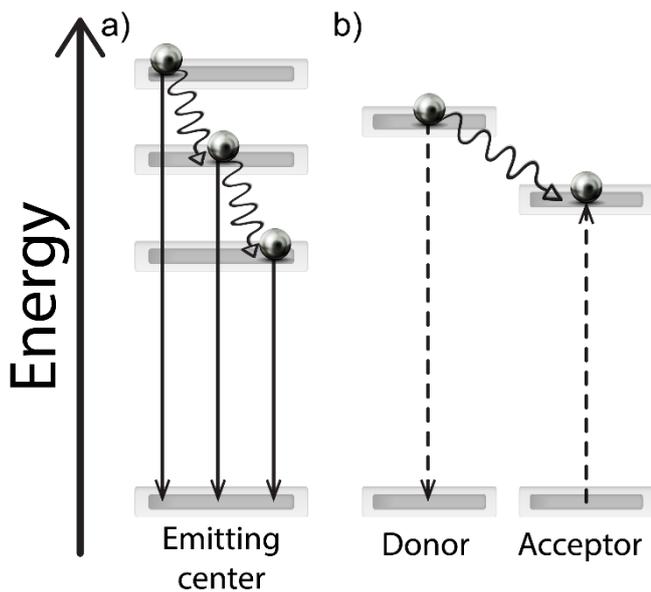

**FIG. 1.**
Representative scheme of non-radiative processes (a) intra-ion multi-phonon relaxation, and (b) inter-ions phonon-assisted energy transfer. Solid-line and wavy- or dashed-line transitions denote radiative and non-radiative (phonon-induced) processes, respectively.

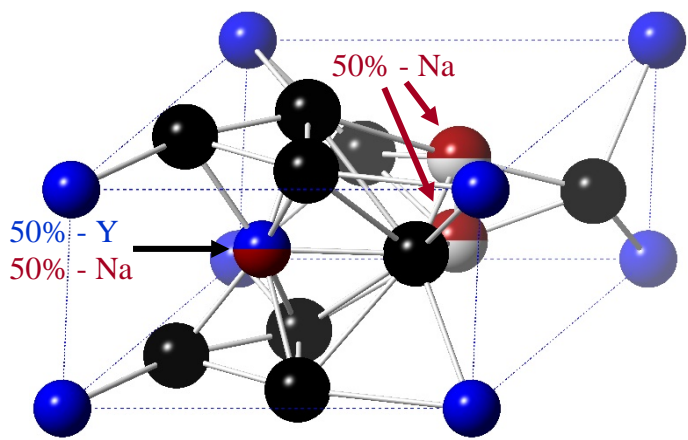

**FIG. 2.**
Hexagonal unit cell for $Na_{1.5}Y_{1.5}F_6$ (#174, $a = b = 5.915$ Å, $c = 3.496$ Å) where the sites with partial Y (blue) and Na (red) occupation have been marked graphically.

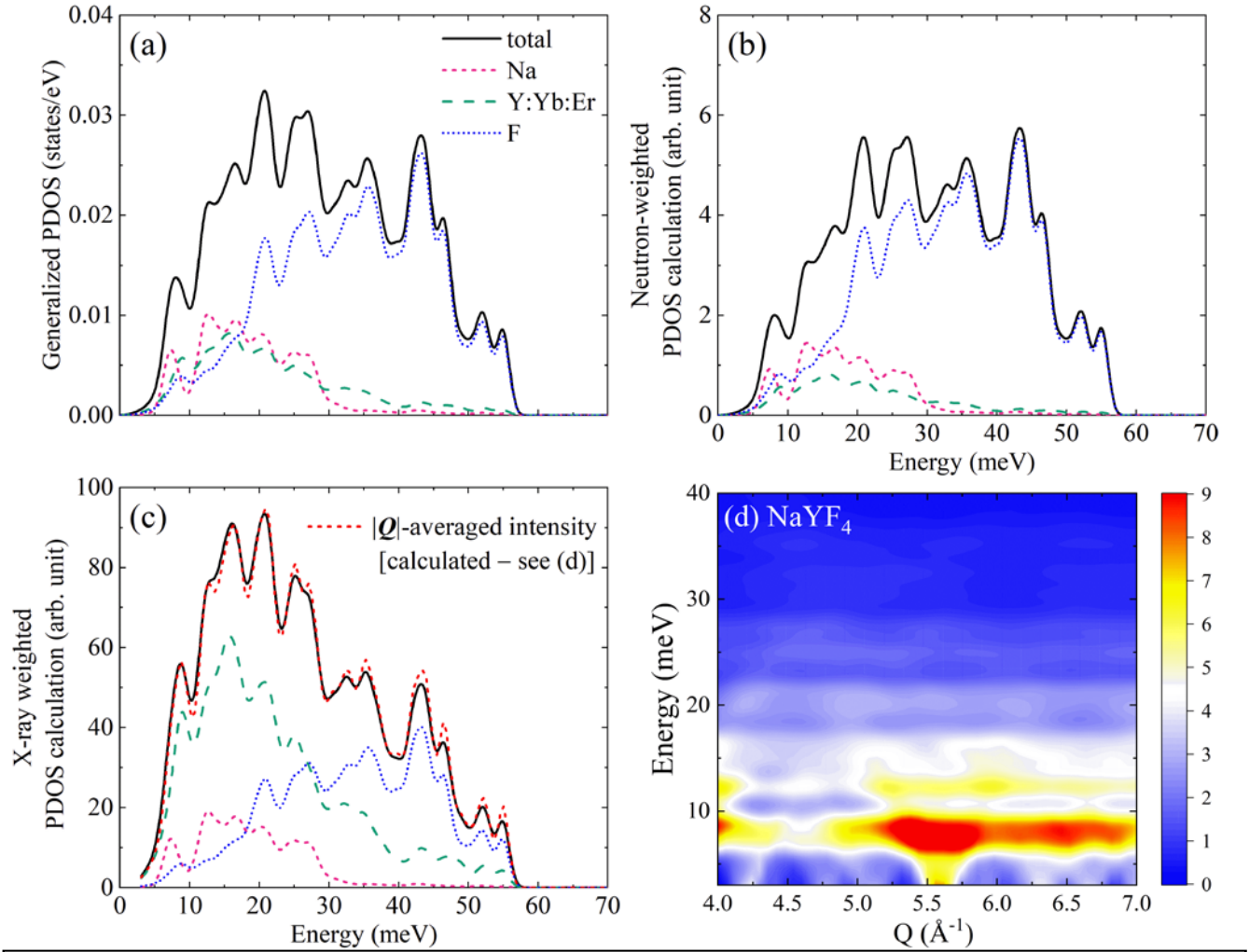

**FIG. 3.**
(a) Calculated generalized PDOS for $NaYF_4$ (solid line) and partial PDOS (broken lines) for Na (purple), Y:Yb:Er (green) and F (blue). The calculation was done for stoichiometric $NaYF_4$. Subsequently, the average atomic mass for $Y_{0.8}Yb_{0.18}Er_{0.02}$ was used to simulate the chemical substitution of the real compound. The legend applies also for (b) and (c). (b) Neutron-weighted PDOS is generated from the partial PDOS for element $k$ multiplied by the factor of $\frac{\sigma_k}{m_k}$ (see text). (c) The X-ray weighted PDOS is generated from the partial PDOSs [see (a)] multiplied by the x-ray scattering cross section. The red-dashed line denotes the corresponding $|\boldsymbol{Q}|$-average of the results shown in panel (d). (d) Color-coded contour map of IXS phonon intensity in $NaYF_4$ for a large range in energy transfers vs. wave vectors $|\boldsymbol{Q}|$. Calculated intensities are directly comparable to background subtracted inelastic IXS data [see Fig. 5(c)] in that they include corrections for the x-ray scattering cross section, the atomic form factor, the $|\boldsymbol{Q}|^2$ dependence of phonon intensities, the Bose factor at room temperature and the factor energy⁻¹. A broadening of 1.7 meV simulates the experimental energy resolution. Intensities at E > 40 meV are close to zero and, therefore, not shown.

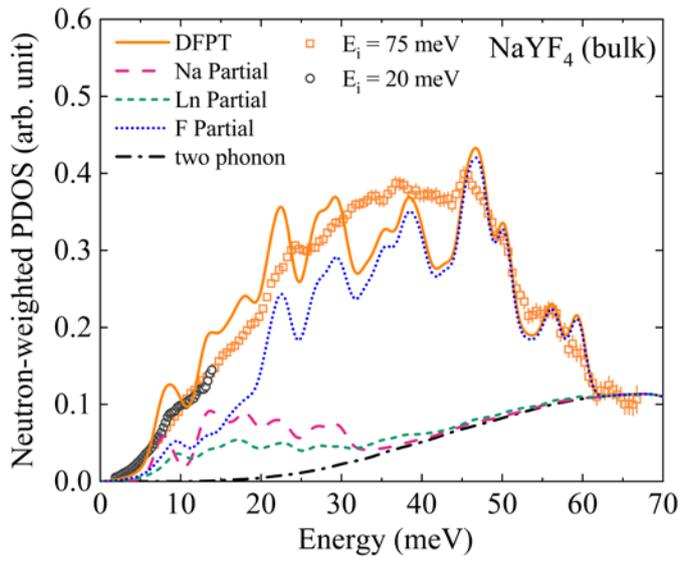

**FIG. 4.**
Neutron-weighted PDOS of $NaYF_4$ obtained with incident neutron energies of $E_i$ = 20 meV (circles) and 75 meV (squares) in comparison to *ab-initio* lattice dynamical calculations (solid line) including the one- and two-phonon PDOS. The partial PDOS are shown as well. The energy scale of the calculation was scaled by a factor of 1.08 and an average broadening of 1.7 meV (FWHM) was applied (see text).

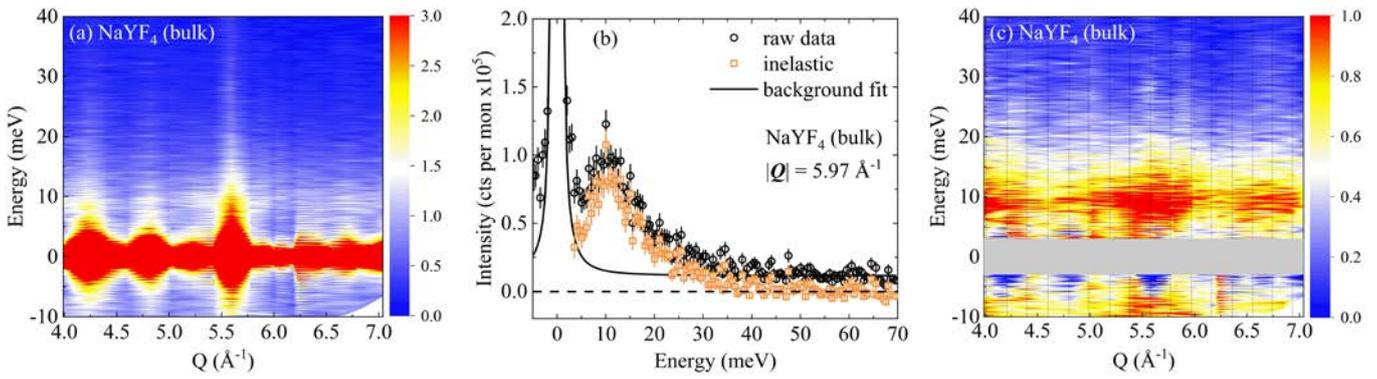

**FIG. 5**
(a) Color-coded contour map of raw IXS data for microcrystalline NaYF₄ over a large range in energy transfer, -10 meV ≤ E ≤ 70 meV, vs. wave vectors |**Q**|, 3.98 Å⁻¹ ≤ |**Q**| ≤ 7.04 Å⁻¹. (b) Raw data at fixed |**Q**| = 5.97 Å⁻¹ (circles). The elastic scattering was approximated by the resolution function (solid line including constant background). Subtraction yields the inelastic spectra (squares). (c) Full energy-momentum range of inelastic spectra taken at 16 different position in |**Q**| (indicated by vertical dotted lines).

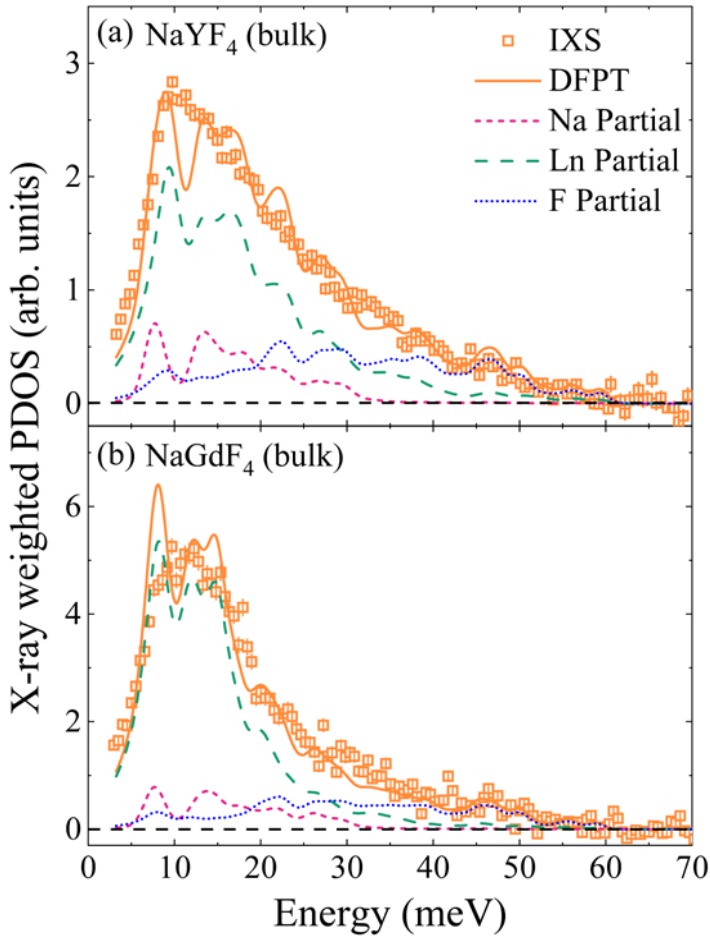

**FIG. 6.**
X-ray weighted PDOS (squares) for (a) NaYF₄ and (b) NaGdF₄ in comparison to corresponding calculations (solid lines). The energy-integrated areas of the calculated results were normalized to the experimental ones. The broken lines show the calculated partial x-ray weighted PDOS of different elements. The energy scale of the calculation was scaled by 1.08, an average broadening of 1.7 meV (FWHM) was applied and the factor 1/E in the phonon cross section was taken into account. Experimental data were corrected for the Bose factor.

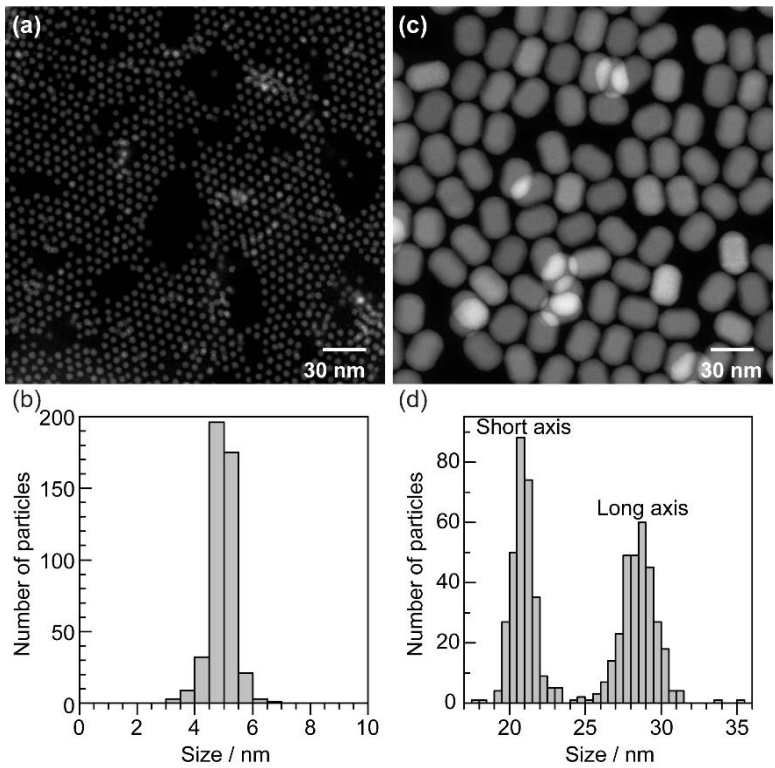

**FIG. 7.**
High-angle annular dark-field scanning transmission electron microscopy (HAADF STEM) images (top) together with their corresponding size distribution histograms (bottom) of (a,b) small and (c,d) large NaGdF$_4$ NCs. Bright regions in (a,c) result from NCs stacked on top of each other.

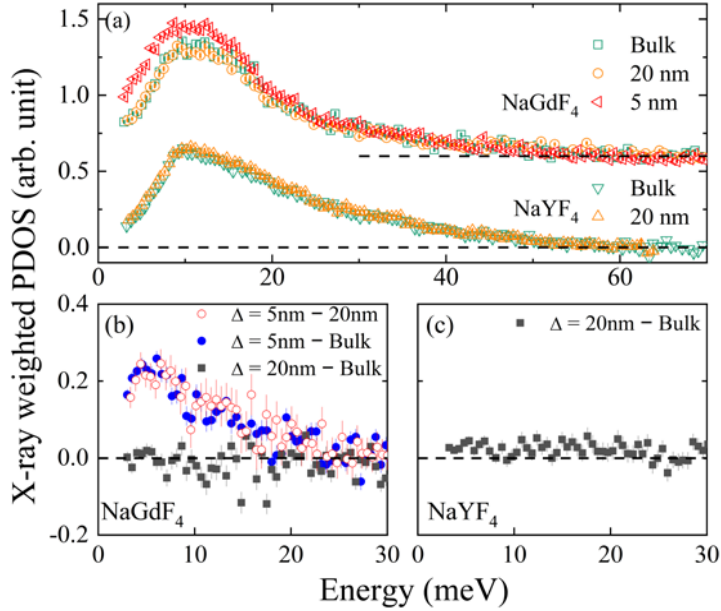

**FIG. 8.**
(a) X-ray weighted PDOS for NaYF$_4$ and NaGdF$_4$ samples with data in green referring to bulk powder samples. All data are corrected for the Bose factor. Data in orange denote results for 20-nm NCs [see Fig. 7(c,d)]. Data for 5-nm NaGdF$_4$ NCs [see Fig. 7(a,b)] are shown in red. Data for NaGdF$_4$ are offset vertically for clarity (zero indicated by the dashed horizontal line). The data were normalized to have the same energy-integrated intensity for $E \geq 20$ meV. (b,c) Difference between X-ray weighted PDOS for nano- and microcrystalline samples for (b) NaGdF$_4$ and (c) NaYF$_4$. For the former we also show the difference between the two nanocrystalline samples.

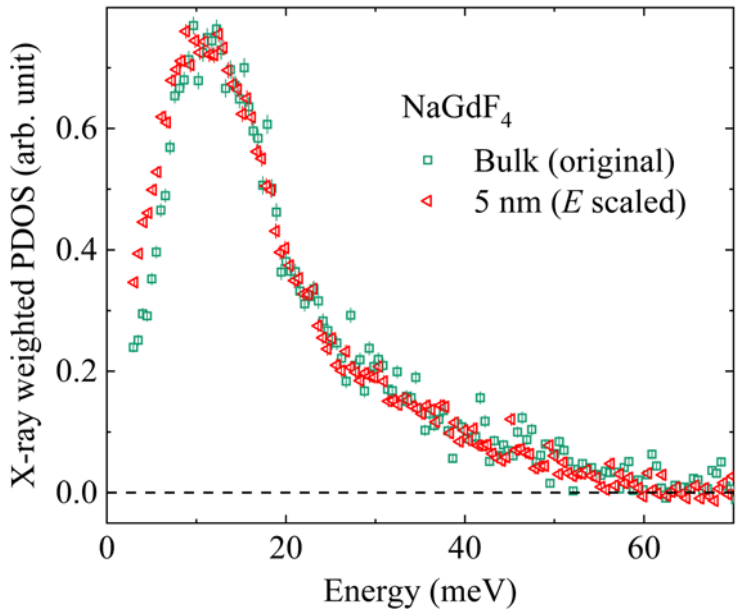

**FIG. 9.**
Comparison of X-ray weighted PDOS for bulk (squares) and 5nm NaGdF$_4$ (triangles) where the energy axis for the latter was scaled by 1.02 in order to compensate the effect of the lattice softening (see text). Data were normalized to the same area.

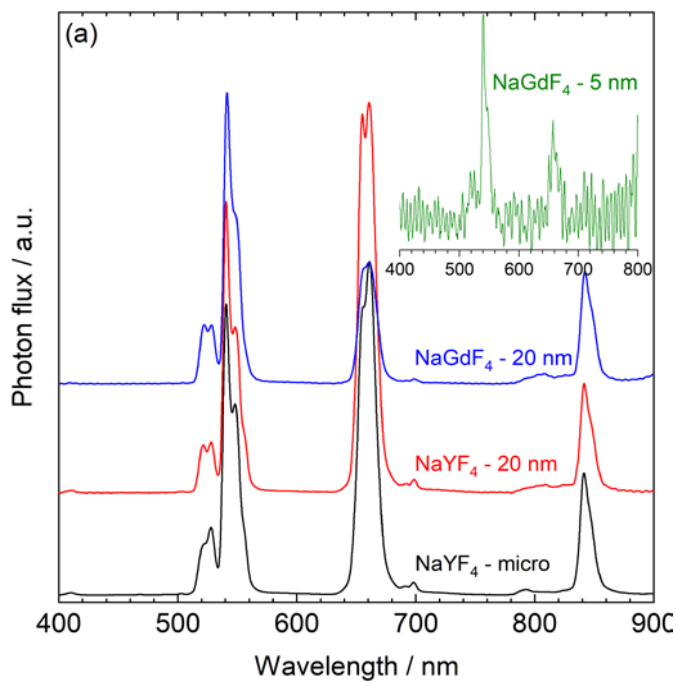

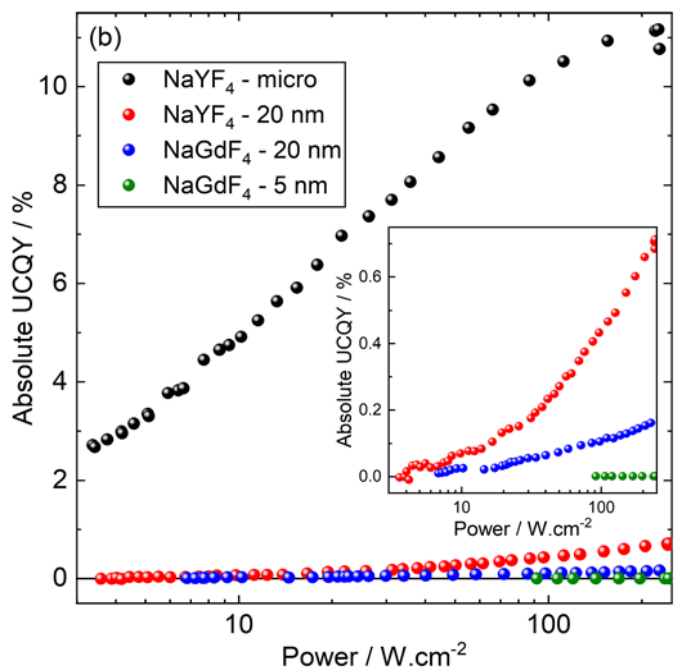

**FIG. 10.**
(a) Emission spectra and (b) power dependent absolute up-conversion quantum yield (UCQY) of microcrystalline NaYF$_4$ (black) compared to large NaYF$_4$ (red) and NaGdF$_4$ nanocrystals (blue) as well as ultra-small NaGdF$_4$ nanocrystals (green).

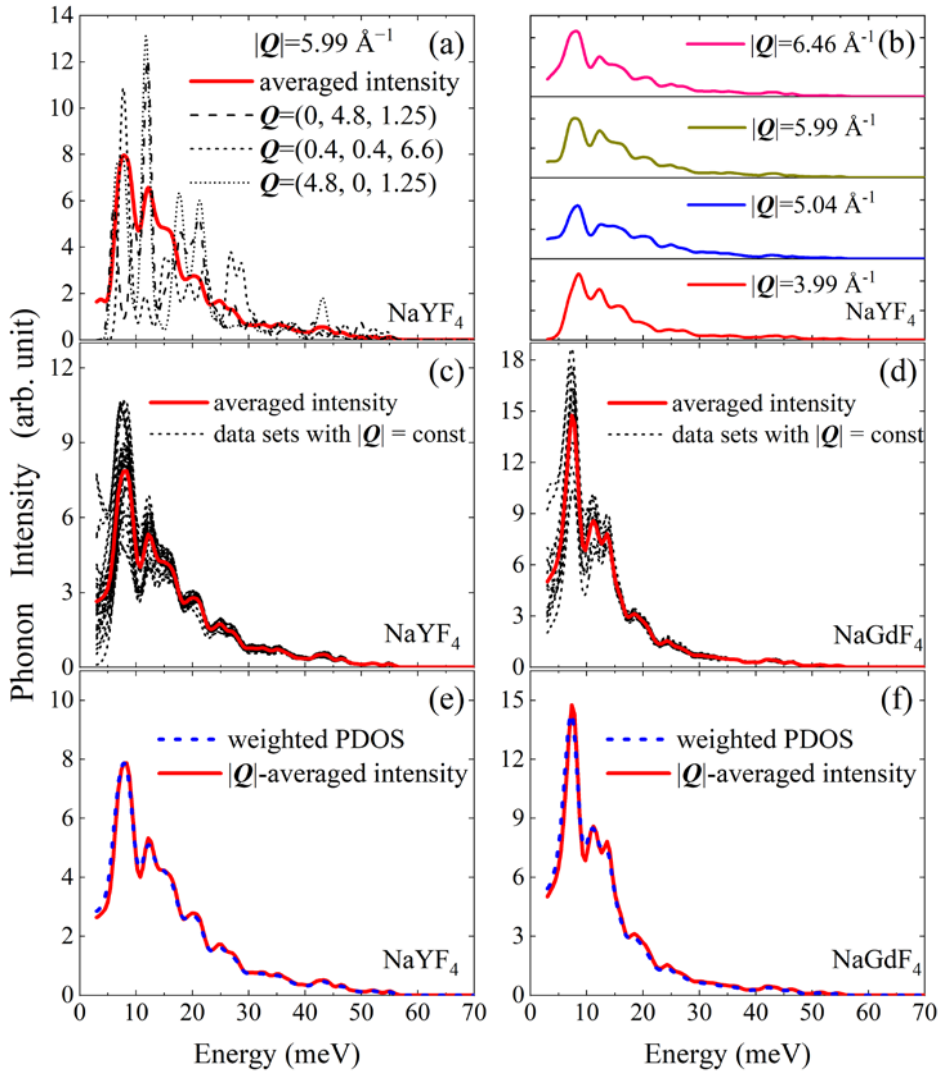

**FIG. 11.**
(a) Calculated phonon intensities at different $\boldsymbol{Q}$ vectors with $|\mathbf{Q}| = 5.99$ Å$^{-1}$ (broken lines) as well as the intensity summed over all calculated $\boldsymbol{Q}$ vectors with the same $|\mathbf{Q}| = 5.99$Å$^{-1}$ and normalized by the number of $\boldsymbol{Q}$ vectors (solid line). (b) Averaged IXS phonon intensities for different values of $|\boldsymbol{Q}|$. (c)(d) Averaged IXS phonon intensities at $|\boldsymbol{Q}|$ values corresponding to the experimentally investigated ones (dotted lines) and the $|\boldsymbol{Q}|$ averaged ones for bulk (c) NaYF$_4$ and (d) NaGdF$_4$. (e)(f) Comparison of the $|\boldsymbol{Q}|$ averaged phonon intensities (red lines, same as in middle row) and the x-ray weighted PDOS for bulk (e) NaYF$_4$ and (f) NaGdF$_4$ (dotted lines). Note that all results in this figure include the Bose factor.

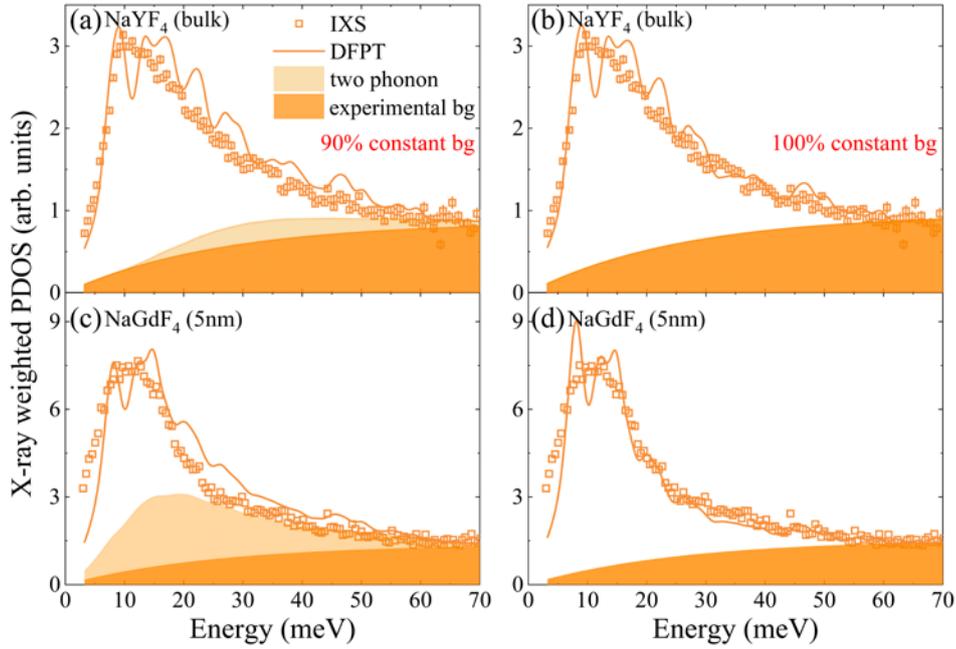

**FIG. 12.**
X-ray weighted PDOS (squares) for (a,b) microcrystalline $NaYF_4$ and (c,d) ultra-small $NaGdF_4$ NCs including the constant background, i.e. only the resolution-limited elastic line was subtracted from the raw IXS data (differently to results shown in Fig. 6). The finite intensity at $E \geq 65$ meV, i.e. above the limit of the one PDOS determined by INS, is subdivided into an experimental background (dark-orange shaded) and a two-phonon contribution (light-orange shaded) with (a,c) 90% and (b,d) 100% of the intensity at $E \geq 65$ meV assigned to the experimental background. The two-phonon contribution is calculated from the calculated X-ray weighted one PDOS (see Fig. 6) convoluted with itself while the experimental background corresponds to a constant value for the raw IXS data which is here shown including a correction for the thermal occupation factor $n+1$ for comparability. Solid lines are the corresponding one PDOS on top of the combined background scaled to match the low energy peak.

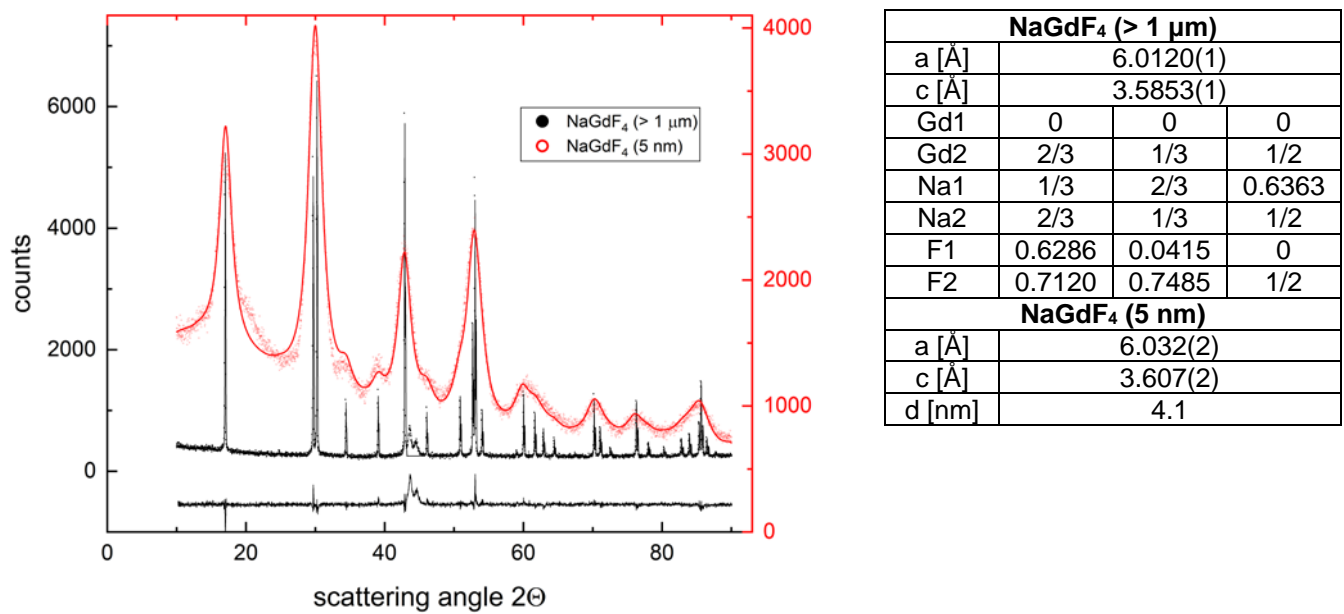

| NaGdF$_4$ (> 1 μm) | | |
|---|---|---|
| a [Å] | 6.0120(1) | |
| c [Å] | 3.5853(1) | |
| Gd1 | 0 | 0 | 0 |
| Gd2 | 2/3 | 1/3 | 1/2 |
| Na1 | 1/3 | 2/3 | 0.6363 |
| Na2 | 2/3 | 1/3 | 1/2 |
| F1 | 0.6286 | 0.0415 | 0 |
| F2 | 0.7120 | 0.7485 | 1/2 |
| NaGdF$_4$ (5 nm) | | |
| a [Å] | 6.032(2) | |
| c [Å] | 3.607(2) | |
| d [nm] | 4.1 | |

**FIG. 13.**
X-ray diffraction (XRD) patterns of microcrystalline (black dots, left-hand scale) and 5 nm NaGdF$_4$ (red circles, right-hand scale). Lines show the respective Rietveld refinement with the refinement parameters given in the table. Atomic positions for the refinement of 5 nm NaGdF$_4$ were fixed to the values obtained for the microcrystalline compound. The black solid line at negative values (left-hand scale, offset -500) shows the difference between the observed and calculated XRD pattern for microcrystalline NaGdF$_4$. The angular range 43.2° ≤ 2Θ ≤ 45.5° shows the presence of a small secondary phase and has been excluded in the refinement. The refined size of the NCs is $d$ = 4.1 nm and is in reasonable agreement with our results from STEM.